# DIRECTION-OF-ARRIVAL ESTIMATION WITH CONVENTIONAL CO-PRIME ARRAYS USING DEEP LEARNING-BASED PROBABILISTIC BAYESIAN NEURAL NETWORKS

*Wael Elshennawy*

Wael.Elshennawy@Orange.com

**ABSTRACT**

The paper investigates the direction-of-arrival (DOA) estimation of narrow band signals with conventional co-prime arrays by using probabilistic Bayesian neural networks (PBNN). A super resolution DOA estimation method based on Bayesian neural networks and a spatially overcomplete array output formulation overcomes the pre-assumption dependencies of the model-driven DOA estimation methods. The proposed DOA estimation method utilizes a PBNN model to capture both data and model uncertainty. The developed PBNN model is trained to do the mapping from the pseudo-spectrum to the super resolution spectrum. This learning-based method enhances the generalization of untrained scenarios, and it provides robustness to non-ideal conditions, e.g., small angle separation, data scarcity, and imperfect arrays, etc. Simulation results demonstrate the loss curves of the PBNN model and deterministic model. Simulations are carried out to validate the performance of PBNN model compared to a deterministic model of conventional neural networks (CNN).

*Index Terms*— Direction-of-arrival (DOA) estimation, co-prime arrays, Bayesian neural networks, neural networks

## 1. INTRODUCTION

The co-prime arrays are class of sparse arrays, which can achieve higher degrees-of-freedom (DOF) that can be exploited in both beamforming and DOA estimation [1]. The coprime arrangement has shown to possess the capability of cancelling spatial aliasing [2], Though the side lobes may still exist in the beampattern that affects the resolution of a DOA estimation algorithm. Therefore, DOA estimation approaches are needed to further explore the advantage of the co-prime arrays. The earlier approaches rely on the subspace-based DOA estimation methods such as multiple signal classification (MUSIC) [3, 4], etc. Meanwhile, these methods require spatial smoothing to restore the rank of the signal covariance matrix [5]. A short and non-exhaustive list of recent works is based on sparse reconstruction so as to use all the unique lags [2, 6]. However, these model-driven methods face great robustness challenges under non-ideal conditions [4].

Another approach provides robust performance against non-ideal conditions include the use of deep convolutional neural networks in [2, 7]. Nevertheless, it is based on the deterministic neural networks. The necessity to develop an approach that exhibits a robustness to the adverse environment. Probabilistic deep learning removes this limitation by quantifying and processing the uncertainty [8]. To further tackle the model and data uncertainty, an off-grid DOA estimation method is proposed from the perspective of variational Bayesian inference [9]. Motivated by the advantages of Bayesian neural networks in [10], this deep probabilistic model is developed based on the normalizing flows for Bayesian neural network to model complex probability distributions [11].

Generally, existing sparsity-inducing DOA estimation methods based on sparse Bayesian learning (SBL) have been demonstrated to achieve enhanced precision [7]. However, the learning process of those methods converges much slowly when the SNR is relatively low. To overcome this challenge in this paper, the coprime arrays is used which provides high SNRs. Regarding the PBNN model, it offers adaptation to various array imperfections and enhanced generalization to unseen scenarios. In addition to, PBNN focuses on marginalization, the estimates would be maximum a posteriori (MAP), and it relies on variational inference and normalization flows to find the optimal values. Its goal is to quantify the model and data uncertainty to explain the trustworthiness of the prediction. Thereby avoiding overfitting problem.

The main contribution of this paper is mainly to consider a probabilistic approach integrated with the deep learning that allows to account for the uncertainty in DOAs estimation of co-prime arrays [11], so that the trained model can assign less levels of confidence to incorrect DOAs predictions. The PBNN model is implemented by using the TensorFlow Probability (TFP) library [12]. Many concepts have been used throughout this paper, including latent variables [13], probabilistic layers [14], bijectors [15], evidence lower bound (ELBO) optimization, and Kullback-Leibeier divergence (KL) divergence regularizers [16] to develop the PBNN model. The presented deep learning approach tends to bring more reliable DOAs estimation, and it has the potential to be applied in real-world environments.

## 2. SIGNAL MODEL OF CO-PRIME ARRAYS

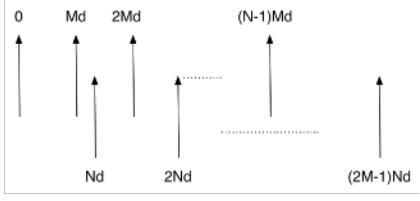

**Fig. 1**: Geometry of co-prime arrays. Adapted from [3].

The co-prime arrays are the union of two uniform linear sub-arrays as illustrated in Fig. 1. One sub-array consists of $2M$-elements with a spacing of $N$ units. The other composed of $N$-elements with a spacing of $M$ units. The positions are given by the set $\mathbb{P}$ in [6] as

$$\mathbb{P} = \{Mnd, 0 \leq n \leq N-1\} \cup \{Nmd, 0 \leq m \leq 2M-1\}. \quad (1)$$

Where $M$ and $N$ are co-prime, and it is assumed that $M < N$. The zeroth sensor positions are collocated, so the co-prime arrays consist of $N + 2M - 1$ elements. The fundamental spacing $d$ usually sets to a half-wavelength to avoid the spatial aliasing. $K$ independent narrow band sources $\mathbf{s}(t) = [s_1(t) s_2(t) ... s_K(t)]$ are impinging on the co-prime arrays from the directions $\{\theta_1, ..., \theta_K\}$. The array output is formulated in [5] as

$$\mathbf{x}(t) = \sum_{k=1}^{K} \mathbf{a}(\theta_k) s_k(t) + \mathbf{n}(t) = \mathbf{A}\mathbf{s}(t) + \mathbf{n}(t), \quad (2)$$

where $\mathbf{A} = [\mathbf{a}(\theta_1), \mathbf{a}(\theta_2), ..., \mathbf{a}(\theta_K)]$ denotes the array manifold matrix, and

$$\mathbf{a}(\theta_k) = [e^{-j2\pi d_1/\lambda \sin\theta_k}, ..., e^{-j2\pi d_{N+2M-1}/\lambda \sin\theta_k}]^T \quad (3)$$

is the steering vector corresponding to $\theta_k$. The $d_1, d_2, ..., d_{N+2M-1}$ hold the information of the sparse elements positions. Whereas $[.]^T$ denotes the transpose of a matrix. $\mathbf{s}(t)$ represents the source signals vector with $s_k(t)$ distributed as $\mathcal{CN}(0, \sigma_k^2)$. The source signals are assumed to be temporally uncorrelated. The entries of the noise vector $\mathbf{n}(t)$ are assumed to be independent and identically distributed (i.i.d) random variables. Also, $\mathbf{n}(t)$ follows a complex Gaussian distributed $\mathcal{CN}(0, \sigma_n^2)$, and their entries are not correlated with source signals.

## 3. PROPOSED APPROACH

Many machine learning models, like deep neural networks, are capable of automatically extracting the necessary features from large inputs. However, there are two main problems with this approach i.e., computational complexity and it requires lots of training dataset. The problem is that, as the machine learning model grows, it becomes more complex. It would take a lot of processing power to automatically learn what features are useful for the classifier, and the use of feature extractor with high dimension inputs is challenging in real time. Also, it is required to store at least that amount of data in memory and has to perform mathematical operations on each of those values. Feature extraction sections in such a model can often be several layers deep, in addition to the computational complexity required to perform inference.

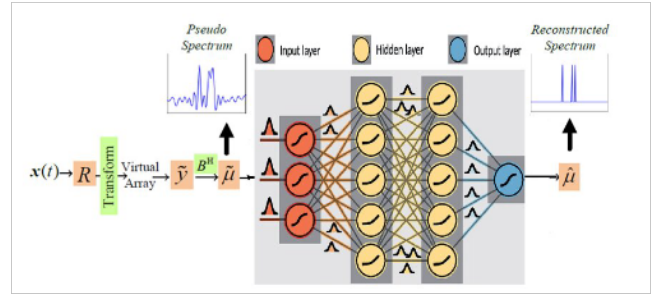

**Fig. 2**: Architecture of the proposed DOA estimation method. Adapted from [2].

Thus, it eventually requires an endless supply of training data and an endless amount of time for training. Though, we do not have that luxury in applications like DOA estimation problem especially it is in the nearest future will be programmed into embedded systems. One ultimately wants to keep machine learning model as small and fast as possible. Most machine learning algorithms require a lot of memory and processing power, and those are in limited supply in most embedded systems. Feature extraction fulfills this requirement: it builds valuable information from raw dataset. The features cycle through reformatting, combining, transforming primary features into new ones, until it yields a new set of data that can be consumed by the machine learning models to achieve their goals. Also, it includes filters for a much faster alternative, filters usually do not test any algorithm, but rank the original features according to their relationship with the problem (labels) itself and just select the top of them. Here, this is features extractor for the PBNN model is achieved by employing the preprocessing step as outlined in the next subsection.

The proposed DOA estimation method for co-prime arrays is illustrated in Fig. 2. The array output is preprocessed to be used by a Bayesian neural network-based model for classification. The pseudo spectrum is calculated from the observation vector and the extended array manifold matrix of a virtual array. This pseudo spectrum is used as the input vector of the PBNN model, and the corresponding super resolution spectrum will be recovered in the output. Thus, this allows to integrate the probabilistic deep learning into a super-resolution DOA estimation method. In addition to, this processing fully maintains the virtual array and effectively im-proves the original SNR.

## 3.1. Preprocessing and Feature Extraction

The covariance matrix $\mathbf{R}$ is given by [1]

$$\mathbf{R} = \mathbb{E}[\mathbf{x}(t)\mathbf{x}^H(t)] = \sum_{k=1}^{K} \mu_k \mathbf{a}(\theta_k)\mathbf{a}^H(\theta_k) + \sigma_n^2 \mathbf{I}, \quad (4)$$

where $\mathbf{R}$ can only be estimated using $Q$ snapshots in practical applications, i.e.

$$\hat{\mathbf{R}} = \frac{1}{Q}\sum_{q=1}^{Q} \mathbf{x}(t_q)\mathbf{x}^H(t_q) = \mathbf{R} + \Delta\mathbf{R}, \quad (5)$$

where $\hat{\mathbf{R}}$ is the maximum likelihood estimator of $\mathbf{R}$ and $\Delta\mathbf{R}$ is the estimation error of $\mathbf{R}$ [2]. By vectorizing $\hat{\mathbf{R}}$, the observation vector of the virtual array is given in [3] as

$$\begin{aligned}\mathbf{y} &= \mathbf{vec}(\hat{\mathbf{R}}) = \mathbf{vec}(\mathbf{R}) + \mathbf{vec}(\Delta\mathbf{R}) \\ &= \tilde{\mathbf{A}}\boldsymbol{\mu} + \sigma_n^2 \mathbf{vec}(\mathbf{I}) + \Delta\mathbf{y},\end{aligned} \quad (6)$$

where

$$\begin{aligned}\tilde{\mathbf{A}} &= [\mathbf{a}^*(\theta_1) \otimes \mathbf{a}(\theta_1), \mathbf{a}^*(\theta_2) \otimes \mathbf{a}(\theta_2), ..., \mathbf{a}^*(\theta_K) \otimes \mathbf{a}(\theta_K)] \\ &= [\tilde{\mathbf{a}}(\theta_1), \tilde{\mathbf{a}}(\theta_2), ..., \tilde{\mathbf{a}}(\theta_K)],\end{aligned} \quad (7)$$

$\otimes$ represents the Kroncher product and $(.)^*$ is the conjugate operation. The signal of interest becomes $\boldsymbol{\mu} = [\mu_1, \mu_2, ..., \mu_K]^\mathbf{T}$, $\mu_k$ denotes the input signal power of the $k$th sources and $\Delta\mathbf{y} = \mathbf{vec}(\Delta\mathbf{R})$, where $\Delta\mathbf{y}$ becomes negligible as the number of snapshots $Q \to \infty$ under stationary and ergodic assumptions. Note that $\mathbf{y}$ amounts to the received data from a virtual array with a much larger aperture defined by the virtual steering matrix $\tilde{\mathbf{A}}$ having the co-array lag locations [5]. Therefore $\tilde{\mathbf{A}}$ behaves like the manifold of a longer equivalent virtual array [6].

Next, by removing the repeated elements of $\mathbf{y}$ and sorting the remaining in an increasing order from $-MNd$ to $MNd$, the output $\tilde{\mathbf{y}}$ is extracted without redundancy for a linear model [3]. By extending the corresponding steering vector, the output of the virtual array can be reconstructed in [1] as

$$\tilde{\mathbf{y}} = \mathbf{B}\boldsymbol{\mu} + \sigma_n^2 \mathbf{vec}(\mathbf{I}),$$
$$\mathbf{B} = [\mathbf{b}^*(\theta_1) \otimes \mathbf{b}(\theta_1), \mathbf{b}^*(\theta_2) \otimes \mathbf{b}(\theta_2), ..., \mathbf{b}^*(\theta_W) \otimes \mathbf{b}(\theta_W)], \quad (8)$$

where $\mathbf{B} \in \mathbb{C}^{(N+2M-1)^2 \times W}$. $\boldsymbol{\mu} = [\mu_1, \mu_2, ..., \mu_W]^\mathbf{T}$, $W \geq K$. $[\theta_1, \theta_2, ..., \theta_W]$ is sampled from the spatial spectrum of incident signals with an interval of $\Delta\theta$. The spatial spectrum $\boldsymbol{\mu}$ is constructed with $W$ grids, which has nonzero values at the true signal directions. The pseudo-spectrum is given by [1]

$$\tilde{\boldsymbol{\mu}} = \mathbf{B}^\mathbf{H}\tilde{\mathbf{y}}, \quad (9)$$

as the input of the Bayesian neural network. This strategy maintains the virtual array generating from co-prime arrays, and it effectively improves the original SNR [7]. To demonstrate the resolution of the pseudo-spectrum $\tilde{\boldsymbol{\mu}}$

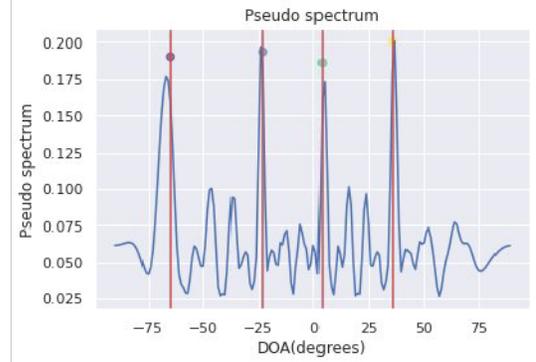

(a) Scenario $A_s$, 4 signal sources with 0dB.

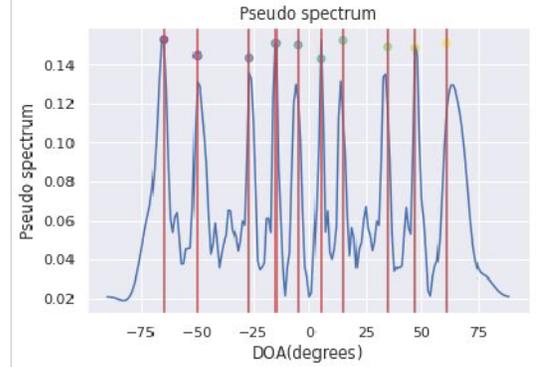

(b) Scenario $B_s$, 10 signal sources with 0dB.

**Fig. 3**: Pseudo spectrum.

of co-prime arrays and shows how it helps in training of the Bayesian neural network. Consider co-prime arrays consisting of 10 physical antenna elements, which is designed by assuming $M = 3$ and $N = 5$. Suppose two different signal sources are impinging on the array from directions sets $A_s = \{-65^o, -23^o, 4^o, 36^o\}$ and $B_s = \{-65^o, -50^o, -27^o, -15^o, -5^o, 5^o, 15^o, 35^o, 47^o, 61^o\}$. Clearly, the co-prime arrays can achieve higher DOF and resolution as illustrated in Fig. 3,

## 4. DOA ESTIMATION BASED ON PBNN MODEL

The idea is to create a neural network with weight uncertainty by combining the neural network with Bayesian inference. Usually, there are two categories of uncertainty; aleatoric and epistemic [11], so there is a necessity to introduce a method for designing a deep learning model that accounts for the uncertainty. In practice, and especially considering the dataset as being finite, there will most likely be many possible parameters values that can do a good job of modeling the re-lationship between the dataset inputs and the targets values. If more dataset is being collected, then the model would have more information about that relationship, and the likely sets of model parameters would probably narrow down. This likely set of parameters values given a dataset is represented as a distribution over all possible parameter values and is called the posterior distribution [16]. Conventionally the term weights will be used to refer to weights and biases for the remainder sections of the paper. Here, the PBNN model is based on the use of probabilistic neural network [15] and the probabilistic layers are implemented by employing TFP library [17].

### 4.1. Bayesian Inference and Posterior Probability

The Bayesian approach is usually implemented by using Backprop algorithm [13], that uses variational inference to give an approximation of the posterior distribution over the model weights [9]. Concisely, the true labels and the likelihood function are used to find the best weights of the Bayesian neural network [12]. For instance, the neural network is a function that maps a pseudo-spectrum data point $\tilde{\mu}_i$ to the proper parameters of some distribution. The PBNN model with weights $\mathbf{W}$ is developed to classify data points $\tilde{\mu}_i$. Hence, the neural network prediction (the feed-forward value) $\hat{\mu}_i$ is defined in [15] as

$$\hat{\mu}_i = \text{BNN}(\tilde{\mu}_i|\mathbf{W}). \qquad (10)$$

Determining $\mathbf{W}$ implies that training a model and assuming that the prediction $\hat{\mu}_i$ forms a part of a distribution that the true label is drawn from. Let the data be $\tilde{\mu}_i$ and the true labels $\mu_i$ for $i = 1, ..., N_s$, where $N_s$ is the number of training samples. Then the training dataset is given as

$$\boldsymbol{D} = \{(\tilde{\mu}_i, \mu_i), ..., ((\tilde{\mu}_{N_s}, \mu_{N_s})\}. \qquad (11)$$

For each point $\tilde{\mu}_i$ has the corresponding prediction $\hat{\mu}_i$, where it assumes specifying a distribution in addition to the true label $\mu_i$. The weights of the trained neural network are then those that minimise the negative log-likelihood loss function in [10] as

$$\begin{aligned}\mathbf{W}^* &= \underset{\mathbf{W}}{\arg\min}(-\sum_{i}^{N_s} \log L(\mu_i|\hat{\mu}_i)), \\ &= \underset{\mathbf{W}}{\arg\min}(-\sum_{i}^{N_s} \log L(\mu_i|\text{BNN}(\tilde{\mu}_i|\mathbf{W}))). \end{aligned} \qquad (12)$$

In practice, determining the true optimum $\mathbf{W}^*$ is not always possible. Instead, an approximated value is sought using optimization algorithms such as root mean squared propagation (RMSProp) or adaptive moment estimation (adam) [11].

## 5. SIMULATIONS RESULTS

In this section, the implementation of both deterministic model and PBNN model are presented. DOAs prediction is modeled as a multi-label classification task [17]. The training, validation, and testing datasets are generated by using Keras generator [17]. The simulations are computed by using Python 3 Google compute engine backend enabling graphics processing unit (GPU) in Google colaboratory notebooks with a mounted drive of a size 12.7 GB.

### 5.1. Simulation Settings and Network Training

Consider co-prime arrays consisting of 10 physical antenna elements, which are designed by taking $M = 3$, $N = 5$. The unit spacing $d$ is chosen to be a half-wavelength. The covariance matrices are computed by using 256 snapshots. The spectrum grid of interest $[-15^o, 15^o]$ is sampled using $1^o$ intervals to form 31 spectrum grid units. The PBNN model and deterministic model are trained using two-signal sources. The simulated signal sources satisfy the far-field narrow-band plane wave conditions. The SNRs of signal sources are generated from a range $[-10, 10]$dB with an 1dB interval.

The models are trained for 10-epochs with a mini-batch size of 32, and the samples set is shuffled at every epoch. The models are fine-tuned using the RMSProp optimizer [17] with a learning rate of 0.05. The total number of training dataset and testing dataset samples is 500 and 100 respectively. With 10% validation dataset off training samples set aside to evaluate the models after each epoch. The architecture of the deterministic model is illustrated in Fig. 4. The deterministic model is stacked by the following Keras layers: Conv1D. BatchNormalization, AveragePooling1D, Flatten, Dropout, and Dense [11].

To build PBNN model, the deterministic model is transformed into a probabilistic model as an intermediate step, by setting the output of the model final layer to a distribution instead of a deterministic tensor. Then this probabilistic model can capture the aleatoric uncertainty on the target DOAs. This is implemented by an addition of a probabilistic layer as a final model layer [14]. Next, turning this probabilistic model into a PBNN model that is designed to capture aleatoric and epistemic uncertainty by changing model layers into reparametrization layers [12] as illustrated in Fig. 4. To further embed an epistemic uncertainty into the model weights by replacing the Conv1D and Dense layers of the deterministic model with Convolution1DReparameterization and DenseVariational layers [12] respectively.

These models are trained using the same conditions for comparison purpose. The loss functions, negative log-likelihood and root mean square error (RMS), are used to measure the DOA estimation performance for each model as illustrated in Fig. 5. The PBNN model provides faster convergence at the early stages and lower training loss values throughout the whole training procedure. Considering that the number of trainable variables of PBNN model is quiet large compared to the deterministic model as tabulated in 1. The training and validation loss curves of PBNN model are almost very close which reveal that the PBNN model is well generalized. Clearly, the validation loss curves level off before 10-epochs. Thus, there is no overfitting in the training phase of the PBNN model.

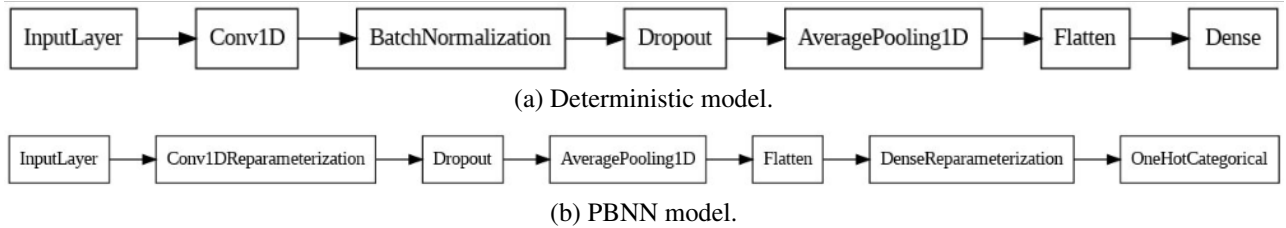

(a) Deterministic model.

(b) PBNN model.

**Fig. 4**: Deterministic model and PBNN model architectures.

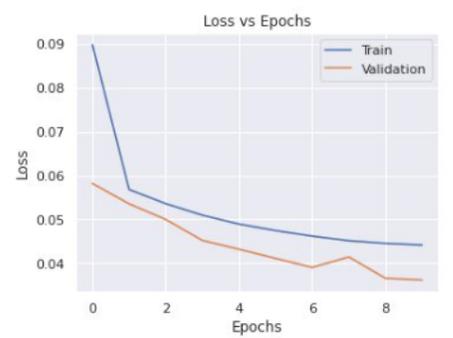

(a) Deterministic model.

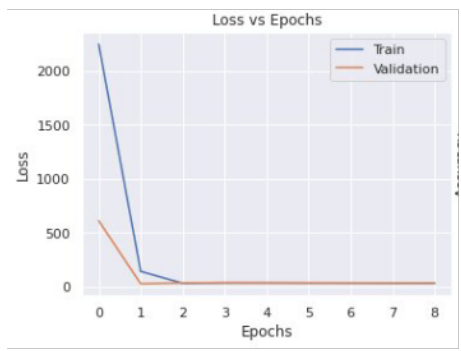

(b) PBNN model.

**Fig. 5**: Training and validation losses versus epoch.

Also, the training step covers different scenarios including changing of the angle of separation, and number of snapshots, etc. Model uncertainty due to insufficient data availability for the model to learn effectively, this is already mitigated by increasing the size of the training data generated by using Keras data generator, which is based on data augmentation method for better model regularization. BNN place a probability distribution on network weights and give a built-in regularization effect making the proposed PBNN model able to learn well from small datasets without overfitting. By introducing a prior, and posterior probabilities, so it preserves the uncertainty that reflects the instability of statistical inference of a small number of instances of evidence dataset. The two properties sparsity of the recovery method and stability are at odds of each other, but the variational Bayesian interference introduces algorithmically stable model.

The PBNN model consists of only 2-hidden network layers as illustrated in Fig. 4. Therefore, it would be useful to increase the size of the model, e.g. by stacking extra network layers for lower loss and MSE values for the case of a wider angular spectral range. Though, the ultimate goal is to develop a simple PBNN model for easy deployment on micro controller chip providing a tinyML operating on the realm of edgeAI. This can serve as an ultra-low power machine learning at the edge.

Finally, testing the angular resolution of the trained PBNN model by incrementally changing the angular separation between two closely spaced signal sources for an angular range varying between $1^o$ and $7^o$ per step size $1^o$. As illustrated in Fig.6, It is obvious that the PBNN model indeed learned to predict the DOAs, and the PBNN model shows robustness. Most importantly, probabilistic Bayesian neural network gives a built-in regularization effect making the PBNN model able to learn well from small datasets without overfitting. Though, Bayesian estimation is computationally very expensive since it greatly widens the parameter space [14]. The pros and cons must be weighed by the user to determine whether the choice of this neural network type is appropriate for the application used. Since the weights of the network are distributions instead of single values, more data is required to accurately estimate the weights.

**Table 1**: Performance comparison between deterministic model and PBNN model.

| Parameter | Deterministic | PBNN |
|---|---|---|
| **Training Loss** | 0.0441 | 33.610 |
| **Validation Loss** | 0.0362 | 31.606 |
| **Training Time (s)** | 1130.0 | 1034.5 |
| **Trainable Variables** | 1,695 | 3,326 |
| **Total Parameter** | 1,727 | 3,326 |

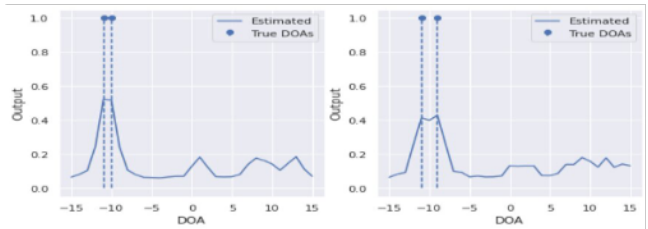

(a) Step-1.

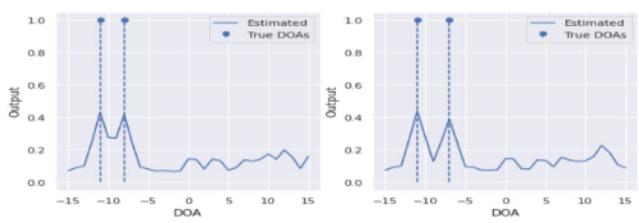

(b) Step-2.

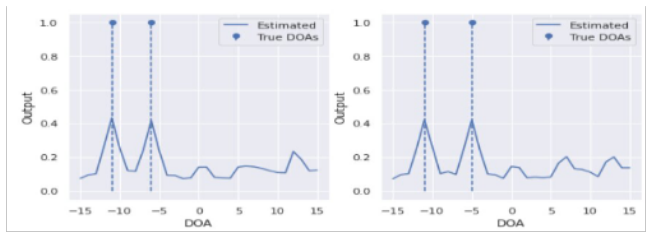

(c) Step-3.

**Fig. 6**: Testing PBNN model by changing the angular separation between two DOAs.

## 6. CONCLUSION

The paper presents a PBNN-based sparse signal recovery method for DOA estimation with co-prime arrays. The DOA estimation based PBNN accounts for the modeling of data and model uncertainty. A CNN is combined with probabilistic layers to learn the mapping from the pseudo-spectrum to the spatial spectrum. The input processing fully maintains the DOF and resolution of the virtual array. The PBNN model can achieve faster convergence at the early training stages and lower training losses. Moreover, the PBNN model adapts well to small angular separation, Simulation results demonstrate that the performance advantages of the PBNN model over deterministic model according to multiple evaluation metrics. Noteworthy, the PBNN model has higher computational complexity than deterministic model. DOAs coarse estimation is obtained by balancing the accuracy and efficiency of parameter estimation using the variational Bayesian-based DOA estimation method. With this PBNN model, the possibility of misclassification is minimized. Thus, this proposed DOA estimation method can achieve spectrum autocalibration under non-ideal conditions for the co-prime arrays. In the future, the goal is to develop the PBNN model for real-time scenario with limited computational resources such as embedded machine learning deployed on a hardware accelerator.

## 7. REFERENCES

[1] Ahsan Raza, Wei Liu, and Qing Shen, "Thinned co-prime arrays for doa estimation," in *2017 25th European Signal Processing Conference (EUSIPCO)*. IEEE, 2017, pp. 395–399.

[2] Ying Chen, Kun-Lai Xiong, and Zhi-Tao Huang, "Robust direction-of-arrival estimation via sparse representation and deep residual convolutional network for co-prime arrays," in *2020 IEEE 3rd International Conference on Electronic Information and Communication Technology (ICEICT)*. IEEE, 2020, pp. 514–519.

[3] Zhao Tan, Yonina C Eldar, and Arye Nehorai, "Direction of arrival estimation using co-prime arrays: A super resolution viewpoint," *IEEE Transactions on Signal Processing*, vol. 62, no. 21, pp. 5565–5576, 2014.

[4] Yimin D Zhang, Moeness G Amin, and Braham Himed, "Sparsity-based doa estimation using co-prime arrays," in *2013 IEEE International Conference on Acoustics, Speech and Signal Processing*. IEEE, 2013, pp. 3967–3971.

[5] Ammar Ahmed, Yimin D Zhang, and Jian-Kang Zhang, "Coprime array design with minimum lag redundancy," in *ICASSP 2019-2019 IEEE International Conference on Acoustics, Speech and Signal Processing (ICASSP)*. IEEE, 2019, pp. 4125–4129.

[6] Zhiguo Shi, Chengwei Zhou, Yujie Gu, Nathan A Goodman, and Fengzhong Qu, "Source estimation using co-prime array: A sparse reconstruction perspective," *IEEE Sensors Journal*, vol. 17, no. 3, pp. 755–765, 2016.

[7] Liuli Wu, Zhang-Meng Liu, and Zhi-Tao Huang, "Deep convolution network for direction of arrival estimation with sparse prior," *IEEE Signal Processing Letters*, vol. 26, no. 11, pp. 1688–1692, 2019.

[8] Moloud Abdar, Farhad Pourpanah, Sadiq Hussain, Dana Rezazadegan, Li Liu, Mohammad Ghavamzadeh, Paul Fieguth, Xiaochun Cao, Abbas Khosravi, U Rajendra Acharya, et al., "A review of uncertainty quantification in deep learning: Techniques, applications and challenges," *Information Fusion*, vol. 76, pp. 243–297, 2021.

[9] Jie Yang, Yixin Yang, and Jieyi Lu, "A variational bayesian strategy for solving the doa estimation problem in sparse array," *Digital Signal Processing*, vol. 90, pp. 28–35, 2019.

[10] Xiao Feng, Xuebo Zhang, Ruiping Song, Junfeng Wang, Haixin Sun, and Hamada Esmaiel, "Direction of arrival estimation under class a modelled noise in shallow water using variational bayesian inference method," *IET Radar, Sonar & Navigation*, vol. 16, no. 9, pp. 1503–1515, 2022.

[11] Stefan Depeweg, *Modeling epistemic and aleatoric uncertainty with Bayesian neural networks and latent variables*, Ph.D. thesis, Technische Universität München, 2019.

[12] Martín Abadi, Paul Barham, Jianmin Chen, Zhifeng Chen, Andy Davis, Jeffrey Dean, Matthieu Devin, Sanjay Ghemawat, Geoffrey Irving, Michael Isard, et al., "Tensorflow: a system for large-scale machine learning.," in *Osdi*. Savannah, GA, USA, 2016, vol. 16, pp. 265–283.

[13] Cheng Zhang, Judith Bütepage, Hedvig Kjellström, and Stephan Mandt, "Advances in variational inference," *IEEE transactions on pattern analysis and machine intelligence*, vol. 41, no. 8, pp. 2008–2026, 2018.

[14] Muhammad Naseer Bajwa, Suleman Khurram, Mohsin Munir, Shoaib Ahmed Siddiqui, Muhammad Imran Malik, Andreas Dengel, and Sheraz Ahmed, "Confident classification using a hybrid between deterministic and probabilistic convolutional neural networks," *IEEE Access*, vol. 8, pp. 115476–115485, 2020.

[15] Oliver Dürr, Beate Sick, and Elvis Murina, *Probabilistic deep learning: With python, keras and tensorflow probability*, Manning Publications, 2020.

[16] Charles Blundell, Julien Cornebise, Koray Kavukcuoglu, and Daan Wierstra, "Weight uncertainty in neural network," in *International conference on machine learning*. PMLR, 2015, pp. 1613–1622.

[17] Orhan Gazi Yalçın and T Istanbul, *Applied Neural Networks with TensorFlow 2: API Oriented Deep Learning with Python*, Springer, 2021.